\documentclass[apj,a4paper,12pt,useAMS]{emulateapj} 
\usepackage{amsmath} 

\usepackage{graphicx,color}

\usepackage{natbib}

\newcommand{\simgt}{\lower.5ex\hbox{$\; \buildrel > \over \sim \;$}}
\newcommand{\simlt}{\lower.5ex\hbox{$\; \buildrel < \over \sim \;$}}

\def\singlebond{\@makechembond\@ne}
\def\doublebond{\@makechembond\tw@}
\def\triplebond{\@makechembond\thr@@}

\shortauthors{Liao et al.}

\shorttitle{Platform deformation correction for AMiBA-13}

\begin{document}

\title{
Platform Deformation Phase Correction for the AMiBA-13 Co-planar Interferometer
}
\author{
Yu-Wei~Liao\altaffilmark{1}, Kai-Yang~Lin\altaffilmark{1}, Yau-De~Huang\altaffilmark{1}, Jiun-Huei~Proty~Wu\altaffilmark{2}, 
Paul~T.~P.~Ho\altaffilmark{1,3}, Ming-Tang~Chen\altaffilmark{1} 
Chih-Wei~Locutus~Huang\altaffilmark{1,2}, Patrick~M.~Koch\altaffilmark{1}, Hiroaki~Nishioka\altaffilmark{1}, 
Tai-An Cheng\altaffilmark{2}, Szu-Yuan Fu\altaffilmark{2}, 
Guo-Chin~Liu\altaffilmark{4}, Sandor~M.~Molnar\altaffilmark{5}, Keiichi~Umetsu\altaffilmark{1}, 
Fu-Cheng~Wang\altaffilmark{2}, Yu-Yen Chang\altaffilmark{6}, Chih-Chiang Han\altaffilmark{1}, 
Chao-Te Li\altaffilmark{1}, Pierre Martin-Cocher\altaffilmark{1}, and Peter Oshiro\altaffilmark{1}
}

\altaffiltext{1}{Institute of Astronomy and Astrophysics, Academia Sinica,
P.~O.~Box 23-141, Taipei 10617, Taiwan; ywliao@asiaa.sinica.edu.tw}
\altaffiltext{2}{Department of Physics, Institute of Astrophysics, \& Center
for Theoretical Sciences, National Taiwan University, Taipei 10617, Taiwan; jhpw@phys.ntu.edu.tw}
\altaffiltext{3}{Harvard-Smithsonian Center for Astrophysics, 60 Garden
Street, Cambridge, MA 02138, USA}
\altaffiltext{4}{Department of Physics, Tamkang University, 251-37 Tamsui,
New Taipei City, Taiwan}
\altaffiltext{5}{Leung Center for Cosmology and Particle Astrophysics, National Taiwan University,
Taipei 10617, Taiwan}
\altaffiltext{6}{Max-Planck-Institut f\"{u}r Astronomie 
   K\"{o}nigstuhl 17. D-69117 Heidelberg, Germany}

\begin{abstract}
We present a new way to solve the platform deformation problem of co-planar interferometers. The platform of a co-planar interferometer can be deformed due to driving forces and gravity. A deformed platform will induce extra components into the geometric delay of each baseline, and change the phases of observed visibilities. The reconstructed images will also be diluted due to the errors of the phases. The platform deformations of The Yuan-Tseh Lee Array for Microwave Background Anisotropy (AMiBA) were modelled based on photogrammetry data with about 20 mount pointing positions. We then used the differential optical pointing error between two optical telescopes to fit the model parameters in the entire horizontal coordinate space. With the platform deformation model, we can predict the errors of the geometric phase delays due to platform deformation with given azimuth and elevation of the targets and calibrators. After correcting the phases of the radio point sources in the AMiBA interferometric data, we recover $50\% - 70\%$ flux loss due to phase errors. This allows us to restore more than $90\%$ of a source flux. The method outlined in this work is not only applicable to the correction of deformation for other co-planar telescopes but also to single dish telescopes with deformation problems. This work also forms the basis of the upcoming science results of AMiBA-13.
\end{abstract}


\section{Introduction}\label{sec:intro}
Co-planar interferometers have been used for cosmological observations for years. Several known co-planar interferometers -- including Degree Angular Scale Interferometer (DASI) \citep{DASI2002}, Cosmic Background Imager (CBI) \citep{CBI2001}, and AMiBA \citep{ho09} -- are dedicated to a variety of topics in cosmology. A co-planar interferometer array benefits from several advantages over an array of individually steerable antennas. Most notably, the co-planar design is good for close-packing a large number of small antennas without the worry of shadowing each other. The servo design of a co-planar array is also simpler because there is no need to control individual antennas simultaneously.

The Yuan-Tseh Lee Array for Microwave Background Anisotropy (AMiBA) is a co-planar radio interferometer for research in cosmology. With an operating frequency band between 86 and 102 GHz, which corresponds to a wavelength $\approx 3$mm, the main goal of AMiBA is to map the Sunyaev-Zeldovich effect (SZE) \citep{SZ1972} in galaxy clusters. The AMiBA antennas and receivers are mounted on a 6-meter carbon fiber platform. The platform itself is mounted on a hexapod mount. The mount allows the platform to be pointed to all the positions in the sky above elevation $30^{\circ}$. The mount can also rotate the platform within a $\pm 30^{\circ}$ range around the normal direction. The polarization angle rotated around the normal direction is called hexpol below.

With 7 platform-mounted 0.6-meter antennas, the former AMiBA-7 observed six galaxy clusters in 2007 and 2008. High quality science data were obtained for these six clusters with AMiBA-7. An overview of AMiBA-7 is given in \cite{ho09}. \cite{Koch2009} detailed the mount and the platform of AMiBA. The stiffness of the AMiBA platform is studied in \cite{Ted2011}. The introductions of correlator and receiver can be found in \cite{Li2006} and \cite{Chen2009}. The details of AMiBA-7 observation, calibration, data analysis, foreground contamination estimation, and quality checking are given in \cite{lin09}, \cite{wu09}, \cite{Liu2010}, and \cite{Nishioka2009}. The science results of AMiBA-7 are shown in \cite{Huang2010}, \cite{Liao2010}, \cite{Liu2010}, \cite{wu09}, and \cite{Umetsu2009}.

\begin{figure*}
\plottwo{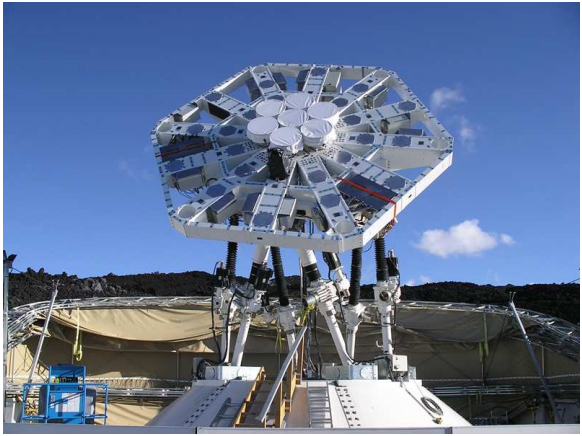}{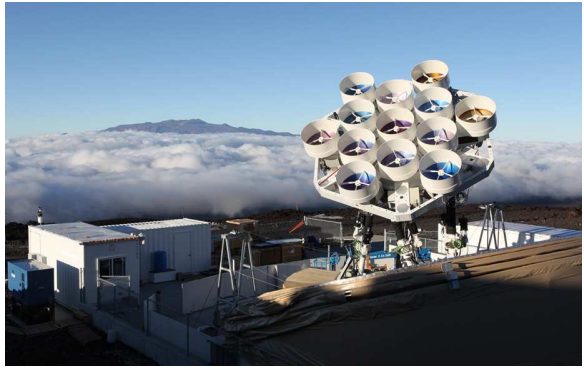}
\caption{The pictures of AMiBA-7 (left) and AMiBA-13 (right). The antennas (0.6m for AMiBA-7 and 1.2m for AMiBA-13) and receivers are mounted in a compact configuration on the 6-meter platform. \label{fig:7and13}}
\end{figure*}

In 2008 and 2009, AMiBA was upgraded to AMiBA-13, which has 13 1.2-meter dishes \citep{Koch2011} mounted on the original AMiBA-7 platform.
The collecting area of AMiBA-13 is 7.5 times of AMiBA-7. The angular resolution of AMiBA was also improved from previously $7'$ to $2'$. With the new AMiBA-13, we can observe fainter objects with a shorter integration time. The ability of AMiBA-13 to constrain intracluster gas models is predicted by \cite{Sandor2010}. Figure \ref{fig:7and13} shows pictures of AMiBA-7 and AMiBA-13.

Driven by the six legs of the hexapod mount, the platform of AMiBA is under complicated pushing and pulling forces from the actuators. The reaction from the actuators, combined with the gravity loading of the interferometer array, cause deformations of the platform. Using photogrammetry, the deformation of the AMiBA platform was measured to be up to several hundred $\mu$m (see \cite{Ted2008} and \cite{Koch2008}). The saddle-shaped platform deformation, which changes its amplitude and direction with different platform pointing and rotation angles, were found to be repeatable in the photogrammetry data. Since we use platform-mounted optical telescopes to calibrate the pointing of the mount, the change of the platform deformations, also induces a change in the relative tilt between the optical telescopes and the platform itself. We found differences in the optical pointing measurement to be $\approx 1'$. \cite{Koch2008} described how the pointing errors have been dealt with in the stage of AMiBA-7.

The other important effect of platform deformations is phase delay. The deformations of the platform will induce extra geometric delays between different antenna elements. In other words, these elements are no longer in the same plane. Those geometric delays will cause artificial phase delays in the observed visibilities from each baseline. 
For AMiBA-7, all the antennas were mounted at the radius $r\leq 1.2$m. Within this range, the platform generally deforms by less than $0.3$mm, or $0.1$ times of the observing wavelength of AMiBA. Therefore, the phase delays are negligible in the AMiBA-7 data. On the other hand, antennas are mounted on the whole 6-m platform for AMiBA-13. The geometric delays between antennas can easily rise to $0.5$ times of the observing wavelength. With this magnitude of geometric delays, the signals are essentially decorrelated, thereby significantly distorting the reconstructed images. Therefore, it is important to find a way to correct the phase delays. The same problem can also be encountered by other co-planer interferometers. Hence our experience of correcting the phase delays might be useful for other telescopes. Furthermore, the way we model and determine the platform deformations can also be applied to study the pointing dependent deformations of single dish telescopes.

\begin{figure*}
\plotone{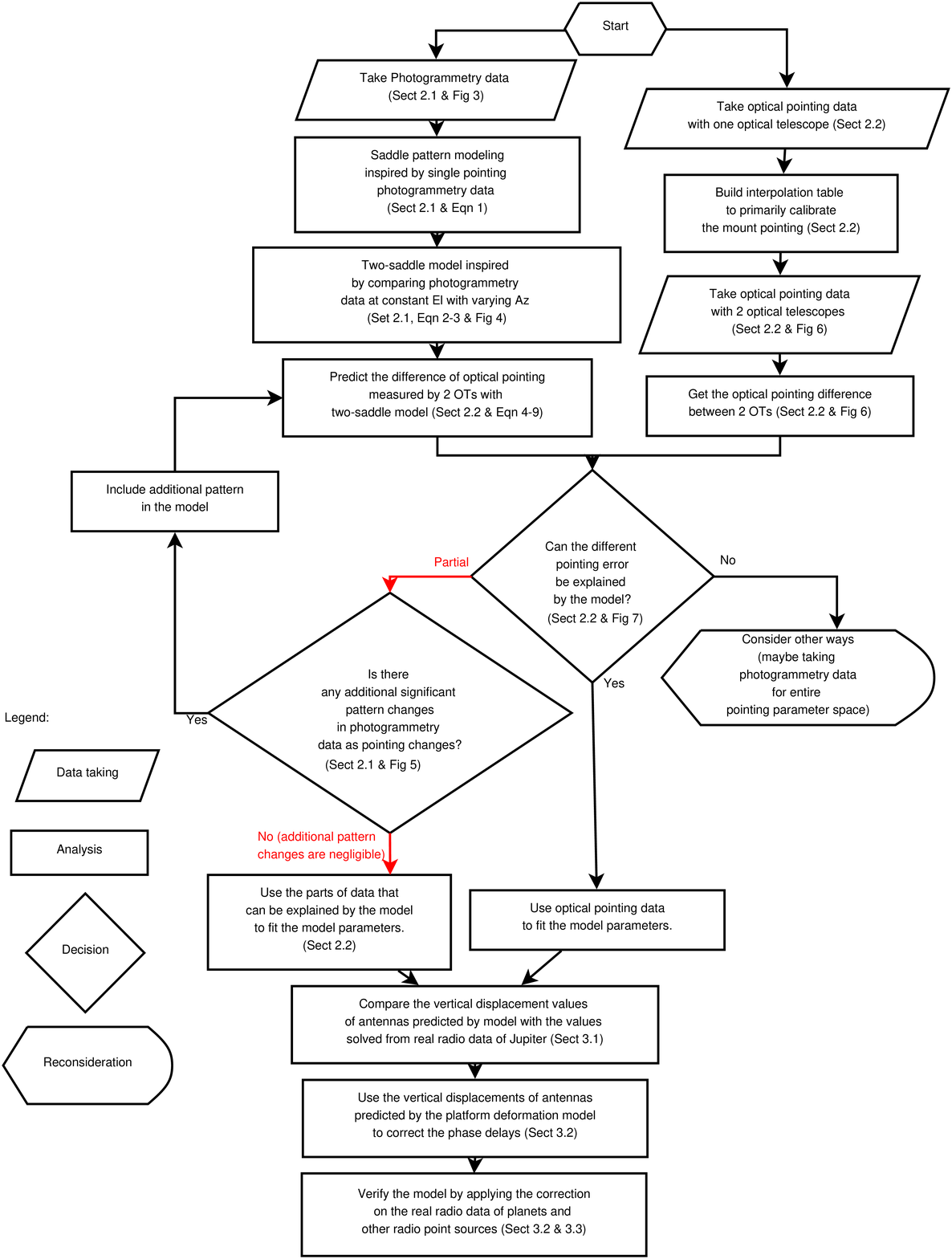}
\caption{The flow chart of our method to correct the platform deformation for AMiBA. The red arrows show the real situations we met. \label{fig:flowchart}}
\end{figure*}

In this paper, we detail the way we model the platform deformations and correct the phase delays (see Figure \ref{fig:flowchart} for the flowchart of our method). In section \ref{sec:platdef}, we describe a new way to model the deformations of the AMiBA platform. We fit the photogrammetry data with a new two-rotating-saddle model. We describe how we correlate the pointing errors of the two platform-mounted optical telescopes to fit the model parameters for the entire pointing parameter space of the mount. In section \ref{sec:corr}, we show how we correct the visibility data base using the platform deformation model. In order to verify the efficiency of our phase correction method, we apply it to the data of planets and radio sources at different pointings on the sky. The results are also shown here.
In section \ref{sec:discuss}, we briefly discuss the improvement of data analysis contributed from the phase correction method, and the limits of such corrections.

\section{Platform Deformation Modeling}\label{sec:platdef}
Before correcting the phase delays contributed by platform deformations, we need to understand more about the phenomenon. The platform deformations of AMiBA are known to be a function of the pointing and the polarization angle of the mount with a repeatability within $50\mu $m (\cite{Ted2008}, \cite{Koch2008}), which is much smaller than the wavelength of $3$mm considered for its application. Therefore, the platform deformations are considered to be modelable and correctable. We describe the photogrammetry-measured platform deformation with an empirical model. However, the photogrammetry surveys were very time-consuming and were, therefore, only done for a sample of selected limited positions. Using two optical telescopes mounted on the platform, it was verified that the relative pointing errors between the two follow the relative tilts of them predicted by the platform deformation patterns. Then we used the relative pointing errors between the two optical telescopes over the whole sky to determine an empirical model of how the deformation pattern changes with pointing.
\begin{figure*}
\begin{center}
\includegraphics[angle=90,height=240mm]{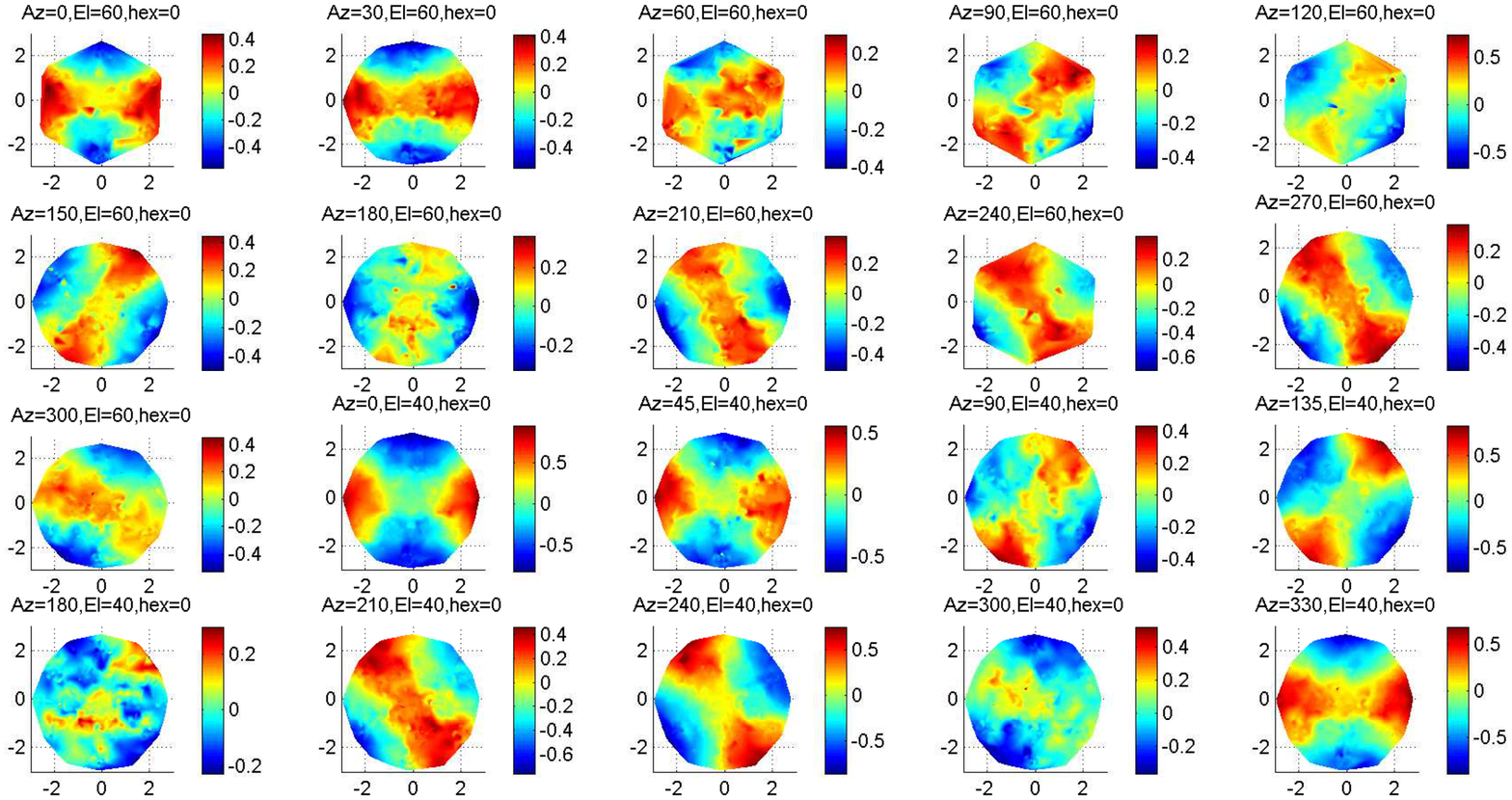}
\end{center}
\caption{Platform deformation. Mount pointing positions (az, el, and hexpol) are indicated in the titles of each panel. The unit of $x$ and $y$ is m, and the unit of $dz$ in color is mm. Obvious saddle features are apparent for most positions. \label{fig:def60}}
\end{figure*}
\begin{figure*}
\plottwo{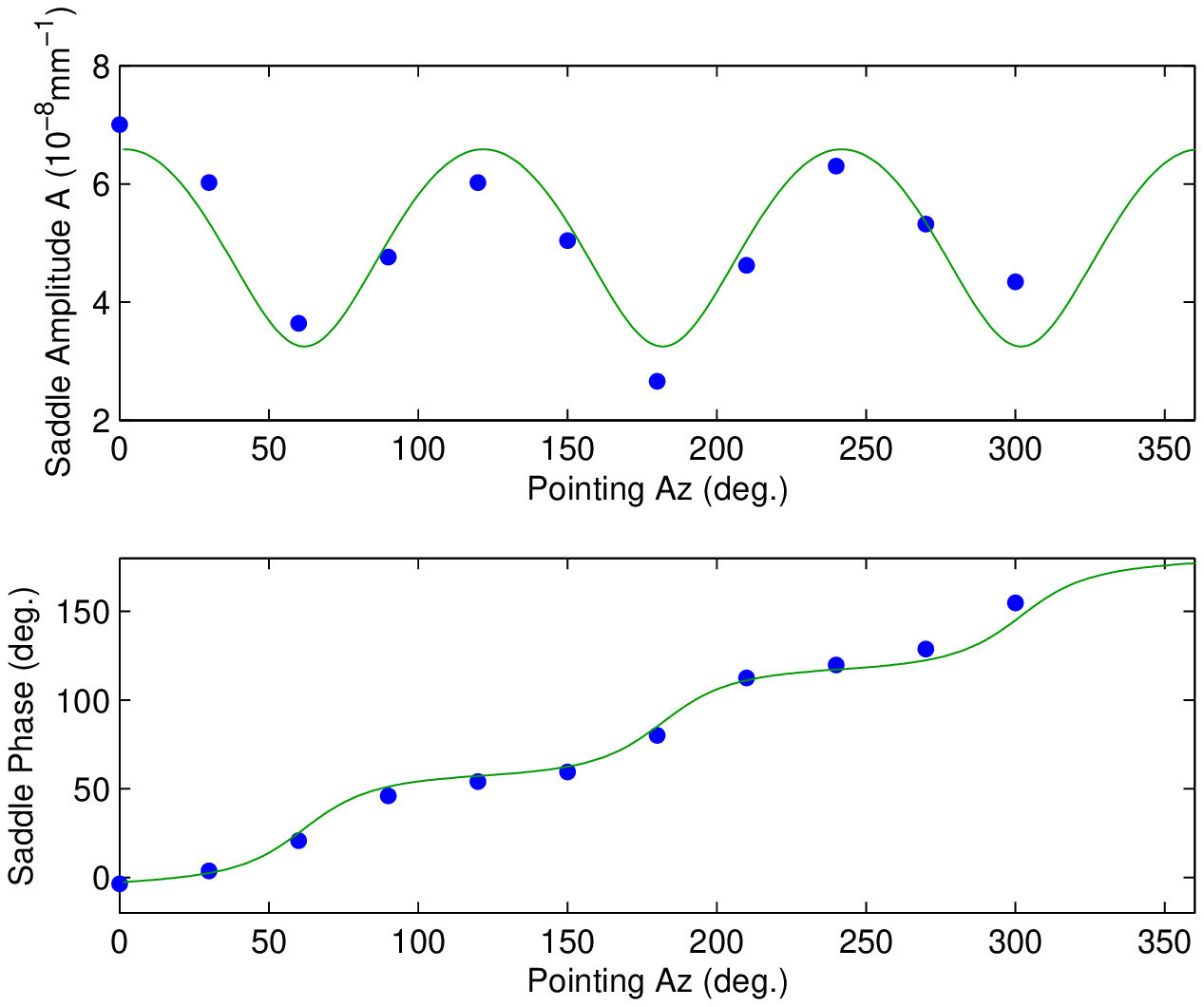}{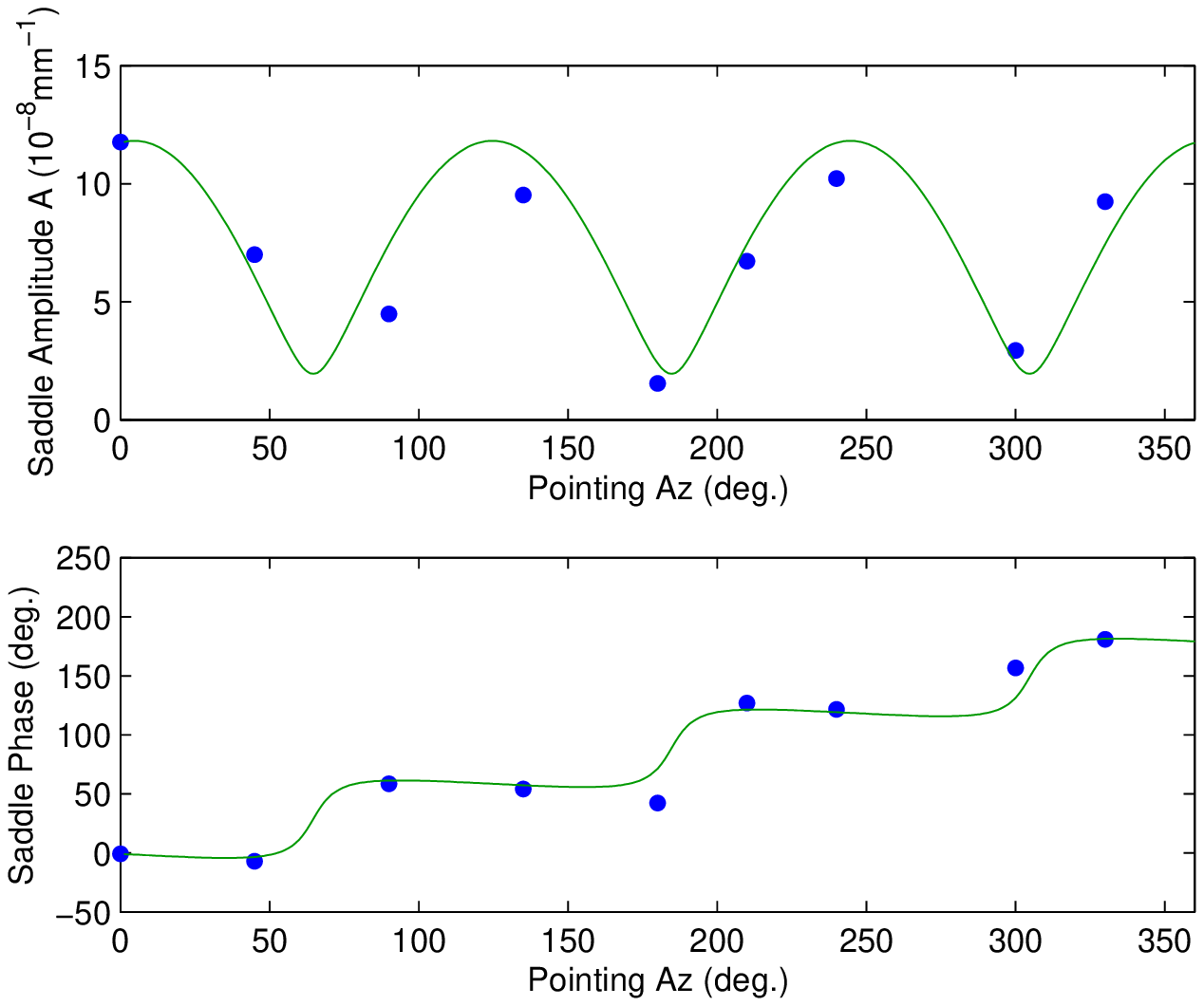}
\caption{Saddle pattern amplitude and phase of platform deformation at $el=60^{\circ}$ (left two panels) and $40^{\circ}$ (right two panels). The hexpol is $0^{\circ}$. The blue dots and green curve show the photogrametric measurements and model prediction, respectively.\label{fig:fit60}}
\end{figure*}
\begin{figure*}
\plotone{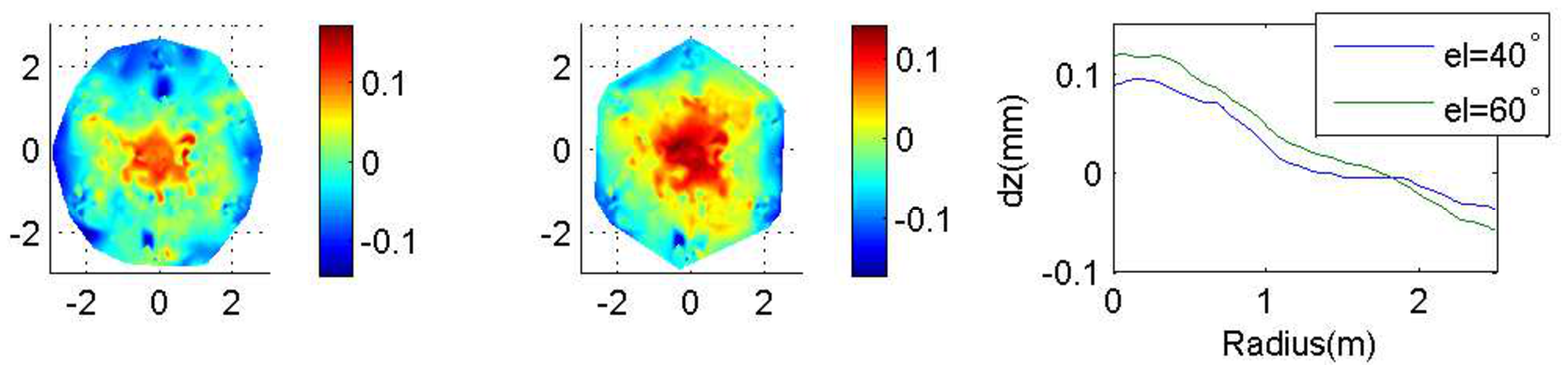}
\caption{The left and middle panels show the residual platform deformation after averaging out all the information dependent on pointing azimuth. The elevation of the left and middle panel are $40^{\circ}$ and $60^{\circ}$, respectively. The unit of $x$ and $y$ is m, and the unit of $dz$ is mm. All data were taken at $hexpol=0^{\circ}$. The right panel shows the azimuthally averaged platform deformation after removing the contribution of pointing azimuth.\label{fig:defresline}}
\end{figure*}
\subsection{Photogrammetry data and deformation modeling}\label{subsec:photo}
Since the loading of the AMiBA platform changed after the upgrade to 13 elements, we did a new photogrammetry test in Nov 2009 (see \cite{Ted2008} for the details of photogrammetry test). The test included 26 sets of data. Each set of data contains several hundreds of photogrammetry measurements taken at different positions on the platform with specified platform pointing and polarization angles. These 26 sets can be split into three groups. The 11 data sets in the first group share the same platform $el=60^{\circ}$ and polarization angle $hexpol=0^{\circ}$. Their platform pointings were distributed uniformly in azimuth between $0^{\circ}$ and $360^{\circ}$. The platform pointings of the 9 data sets in the second group were distributed in the same way in azimuth with $el=40^{\circ}$ and $hexpol=0^{\circ}$. The other 6 data sets were taken at $el=30^{\circ}$, $az=0^{\circ}$, and 6 different polarization positions. Part of the results are shown in Figure \ref{fig:def60}.

The platform deformation can be described as a "saddle pattern", which can be parametrized as following \citep{Koch2008}:
\begin{equation}
dz=A[(x^{2}-y^{2})\cos 2\theta +2xy\sin 2\theta],
\label{eq:saddle}
\end{equation}
where $dz$ is the platform deformation in the normal direction of the platform, $x$ and $y$ indicate the
position on the platform. Here we set east direction as positive $x$, north as positive $y$, and positive $z$ above the platform. The saddle pattern is described by two parameters: $A$, the amplitude of the saddle pattern, and $\theta$, the phase of it.

For each set of photogrammetry data, we can find the best fit $A$ and $\theta$. The best fit parameters of the first two groups of photogrammetry data sets are shown in Figure \ref{fig:fit60} as blue dots. As we can see, the amplitude of the best fit saddle pattern has a 3-fold symmetry. The phase of the saddle pattern changes as $\theta=0.5az+\theta_{0}+\epsilon$, where $\theta_{0}$ is a constant, and $\epsilon$ looks like a periodic perturbation. (Note: because of our definition of $x$ and $y$, the directions of $\theta$ and $az$ are opposite to each other.)

It can be shown that if there are two saddle patterns, $dz_1(x,y)=A_1[(x^2-y^2)\cos 2\theta_1+2xy\sin 2\theta_1]$ and $dz_2(x,y)=A_2[(x^2-y^2)\cos 2\theta_2+2xy\sin 2\theta_2]$, the combination of these two saddle patterns $dz_1+dz_2=A_3[(x^2-y^2)\cos 2\theta_3+2xy\sin 2\theta_3]$ will be another saddle pattern. Furthermore, we can derive that $A_3\cos 2\theta_3=A_1\cos 2\theta_1+A_2\cos 2\theta_2$ and $A_3\sin 2\theta_3=A_1\sin 2\theta_1+A_2\sin 2\theta_2$. In other words, a saddle pattern can be described as a vector $\vec{p}$ with absolute value $A$ and phase angle $2\theta$, and the combinations of saddle patterns can be considered as summations of vectors.

We can consider the best fit saddle pattern parameters shown in Figure \ref{fig:fit60} as the absolute value and half of the phase angle of $\vec{p}$. The 3-fold symmetry of $\vert\vec{p}\vert$ and the periodic perturbation of $angle(\vec{p})$ can be explained as the summation of two vectors $\vec{p}_{1}=(A_{1}\cos{\alpha_{1}},A_{1}\sin{\alpha_{1}})$ and $\vec{p}_{2}=(A_{2}\cos{\alpha_{2}},A_{2}\sin{\alpha_{2}})$ rotating along the opposite directions as $az$ grows. If $\alpha_{1}$ rotates with angular velocity $w_{1}=1$ as the azimuth grows and $\alpha_{2}$ rotates with angular velocity $w_{2}=-2$, $\vec{p}_1$ and $\vec{p}_2$ will be parallel to each other while $az=0,2/3\pi ,4/3\pi$, and anti-parallel while $az=1/3\pi ,\pi ,5/3\pi$. That can explain the 3-fold symmetry of $A$, which corresponds to the absolute value of the summation of $\vec{p}_{1}$ and $\vec{p}_2$. If we assume $A_{1}>A_{2}$, the phase angle $\alpha$ of the combined vector $\vec{p}$ will be swinging around $\alpha_{1}$ with a period of $2/3\pi$. That can also explain what we see in Figure \ref{fig:fit60}.

The two-saddle model described above can be summarized as following:
\begin{eqnarray}
dz&=&A_{1}[(x^{2}-y^{2})\cos 2\theta_1 +2xy\sin 2\theta_1] \nonumber \\
  & &+A_{2}[(x^{2}-y^{2})\cos 2\theta_2 +2xy\sin 2\theta_2],
\label{eq:model2}
\end{eqnarray}
where
\begin{eqnarray}
\theta_1 &=& \frac{az}{2}+\phi_1 \nonumber \\
\theta_2 &=&-az+\phi_2.
\label{eq:model3}
\end{eqnarray}
Equation (\ref{eq:model2}) remains a saddle pattern. 

There are four parameters: $A_1$, $A_2$, $\phi_1$, and $\phi_2$, in the model described by Equation (\ref{eq:model2}). The green curves in Figure \ref{fig:fit60} show $A$ and $\theta$ as predicted by our model with the parameters fitted to the photogrammetry data. We found that our model fits the blue dots well. We also noticed that both $A_1$ and $A_2$ grow significantly as the platform elevation is lowered from $60^{\circ}$ to $40^{\circ}$. But $\phi_1$ and $\phi_2$ do not change much ($<10^{\circ}$) between elevation $40^{\circ}$ and $60^{\circ}$.

There can be additional secondary features of the platform deformation other than the saddle pattern model described above. To investigate this, we subtract the best fit saddle patterns from each set of photogrammetry data, then average the residuals over pointing azimuth. By averaging over pointing azimuth, we can extract the deformation features independent of it. The left and middle panels of Figure \ref{fig:defresline} show the averaged residual data at elevation $40^{\circ}$ and $60^{\circ}$, respectively. We notice that the averaged residual has isotropic features. The residual $dz$ tends to be higher at the center and lower at the edge of the platform. In the right panel of Figure \ref{fig:defresline} we average the residual feature along  concentric circles, which share the same center with the platform itself, on the platform. The difference of $dz$ at the center and the edge is about $0.15$mm, or $0.05\lambda$. The azimuthal average does not change much ($<0.03$)mm as the pointing elevation changes between $40^{\circ}$ and $60^{\circ}$. 

We are now addressing the question whether the secondary features are important to our phase correction work. Considering our calibration method, if the cluster scans and the calibration scans share the constant secondary-feature deformation, the phase delays induced by it will be subtracted during the calibration scheme. Therefore, the key concern is how the platform deformation changes with platform position. This will then determine how to calibrate the data in the absence of a nearby strong calibrator. Based on this consideration, the saddle pattern model described in Equation (\ref{eq:model2}) plays the main role in our platform deformation corrections. Since the secondary deformation pattern only changes within $0.01 \lambda$ as pointing changes from $el=40^{\circ}$ to $el=60^{\circ}$ at $\simeq 95 \% $ of positions on the platform (i.e.: comparing the left and middle panels of Figure \ref{fig:defresline}), it is considered to be negligible in this work. Consequently, secondary features are not further taken into account in our correction scheme.

\subsection{Correlation of 2 Optical Telescopes}\label{subsec:ots}
In principle we can directly measure the platform deformation with photogrammetry measurements. However, it takes time to sample the entire 3-dimensional pointing parameter space ($az$, $el$, and $hexpol$) with acceptable resolution. Fortunately, there is another way using the optical pointing measurement to estimate the platform deformation model for the whole sky.

\begin{figure*}
\plottwo{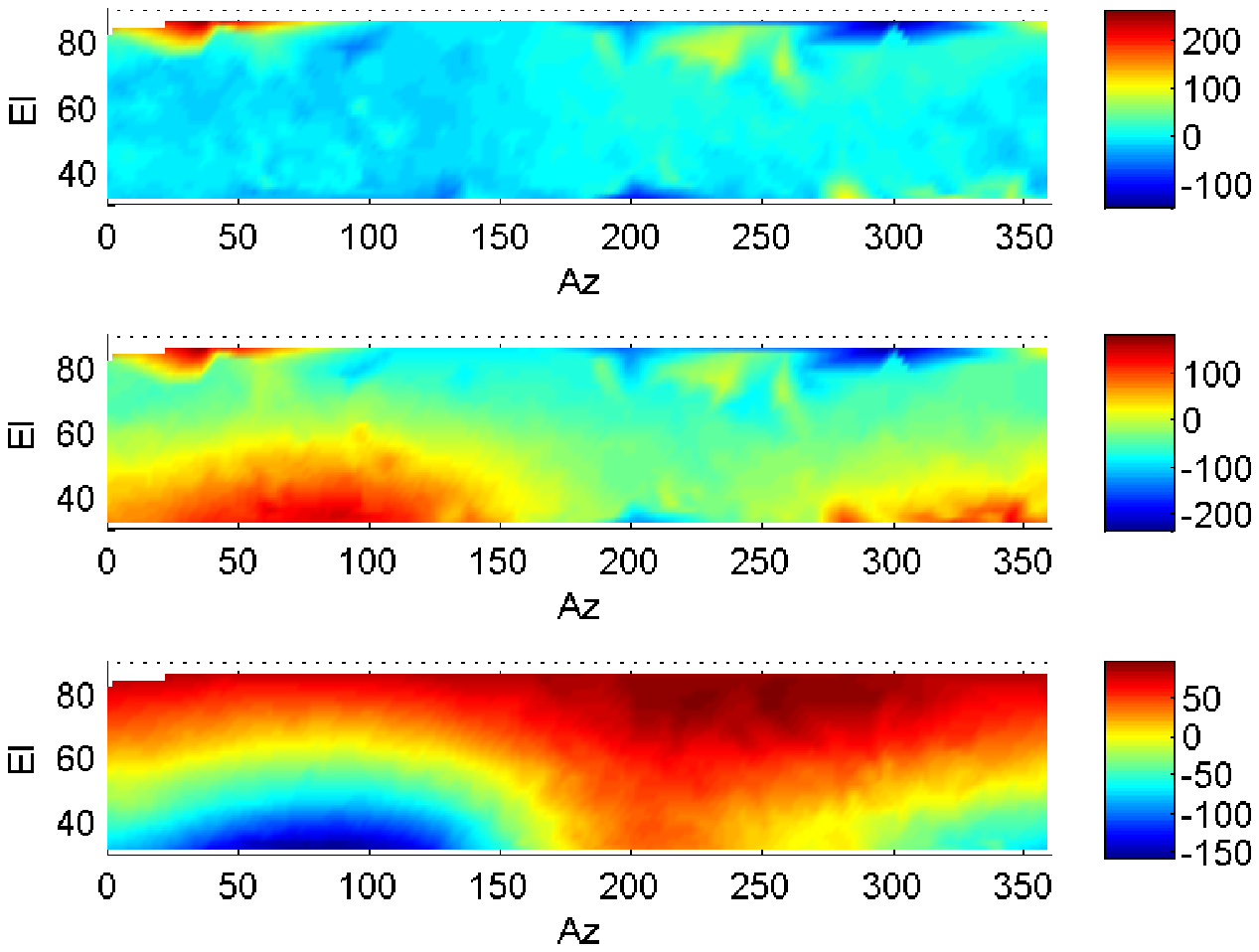}{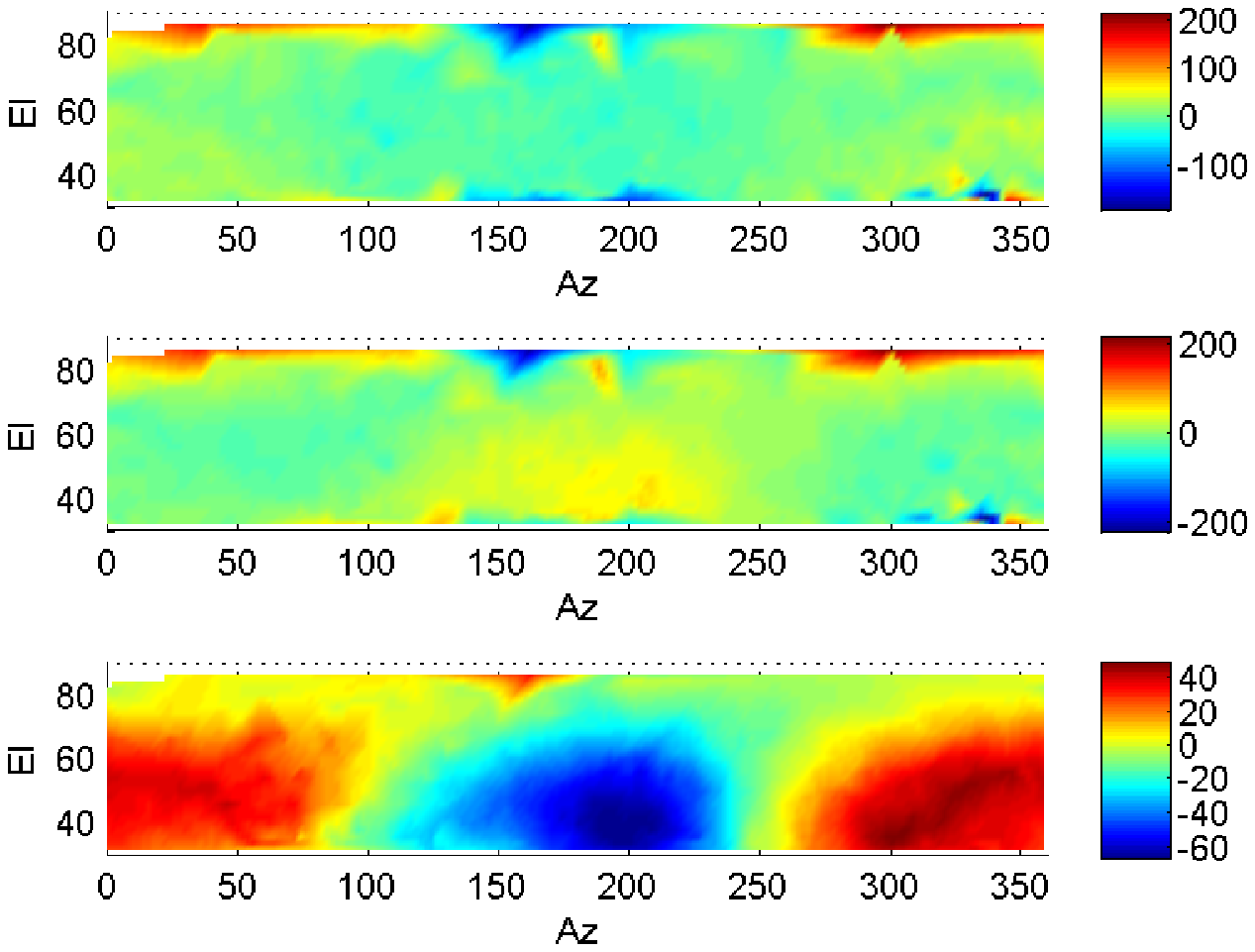}
\caption{The upper left panel is the X-direction pointing error measured by OT1, the middle left panel is the X-direction pointing error measured by OT2, and the lower left panel is the difference between OT1 and OT2 measurement on X-direction pointing error. The right column panels show the Y-direction pointing error in the same order. The unit of $az$ and $el$ are both degree, and the unit of pointing error is arc-second. All the pointing errors in this figure are obtained at $hexpol=0^{\circ}$. The interpolation table built based on OT1 pointing test was applied to correct the pointing error.\label{fig:OTdiff}}

\end{figure*}

\begin{figure}
\plotone{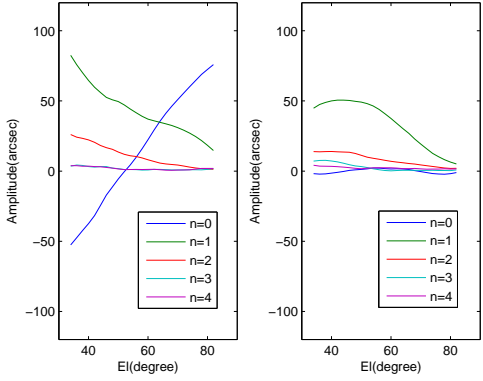}
\caption{Amplitude of Fourier modes obtained by Fourier transforming the difference of pointing errors measured by two OTs (as shown in Figure \ref{fig:OTdiff}) along iso-elevation bands. The left panel shows the results obtained with X-direction pointing error, and the right panel shows the Y-direction results.\label{fig:OTft}}
\end{figure}

There are two optical telescopes (OTs) mounted on the AMiBA platform to test the pointing of the mount.
Because the pointing of a single optical telescope can be significantly affected by the platform deformation
and other local effects, we use two of them to do cross checking. The first optical telescope (OT1) is mounted
at the distance of about 0.6 meters away from the center point of the platform. The other optical telescope (OT2) is mounted at the same radius, but $120^{\circ}$ apart from OT1.

The main idea of the optical pointing test is that the difference of pointing errors taken by two OTs provides information of the relative tilt between two OTs. We can then use the information of this relative tilt to fit the platform deformation model described above. Comparing with the several hundreds of samples provided by the several hundreds of targets on the platform in the photogrammetry test for a single pointing position, the sampling for a single pointing position provided by optical pointing is really poor (only 2 samples for a pointing position). However, if the platform deformation itself can be described well by a simple model like our two-saddle pattern model, the poor sampling on the platform will not cause any severe problem. On the other hand, the optical pointing test can provide $\sim 500$ samples per night in the parameter space of platform pointing and polarization angle. With photogrammetry measurements it would take 2 months to achieve the same sampling level. Since we need to build all-sky phase corrections, optical pointing tests are more efficient and suitable for this goal.

Our optical pointing tests for AMiBA-13 were conducted in 2010. We followed a similar procedure as described in \cite{Koch2008}. The sky was partitioned into 500 zones with equal solid angle above elevation $=30^{\circ}$. One bright star per zone was chosen. We took images at $hexpol=-24^{\circ}, -12^{\circ}, 0^{\circ}, 12^{\circ},$ and $24^{\circ}$ for each star with 2 OTs. Therefore, we have 10 images for each star. It took 5 nights to finish the optical pointing observations. Then we grouped those 2500 images according to OTs and $hexpol$. With each image we derived the pointing error at a specified pointing coordinate. Therefore, we have 500 data points of pointing error with each combination of specified OT and polarization. After removing the constant tilt of the OTs \citep{Koch2008}, we interpolated a pointing error map (e.g.: Figure \ref{fig:OTdiff}) at different hexpol for each OT.

First, we use OT1 to construct an interpolation table (IT), which is used to correct the mount pointing error.
The rms of OT1 pointing error is about $0.3'$ with IT correction. However, the difference of pointing between OT1 and OT2
is at the level of $1'-2'$. Figure \ref{fig:OTdiff} shows the pointing errors of both OTs and the differences between the errors. In Figure \ref{fig:OTdiff} we have transformed the pointing errors into platform coordinates. 

The differences of pointing errors in $X$ and $Y$ directions can be expressed as $perr_{diff,x}(az,el)$ and $perr_{diff,y}(az,el)$, respectively. We fix the elevation to extract
\begin{eqnarray}
g_{x,el_0}(az)=perr_{diff,x}(az,el=el_0) \nonumber \\
g_{y,el_0}(az)=perr_{diff,y}(az,el=el_0).
\label{eq:gxy}
\end{eqnarray}
We then perform Fourier transformations on $g_{x}$ and $g_{y}$ with respect to $az$. Finally, we get Fourier modes
as functions of elevation.

Figure \ref{fig:OTft} shows the Fourier transformation results of the relative pointing errors between 2 OTs. As shown in Figure \ref{fig:OTft}, the dominating Fourier modes of the pointing differences are the modes with $n=0,1,2$. The higher order modes are negligible.

Ideally, the relative pointing errors between OTs can be explained by the platform deformation. Here, we assume that the pointing errors measured by the optical telescopes are the combination of the mount pointing error and the relative tilt between OT and platform. We also assume that the relative tilt between OT and platform is contributed by two origins: generic tilting of the OT box (assumed to be constant) and the platform deformation (assumed to be variable). We adopted the way described in \citet{Koch2008} to remove the constant tilt. In this section we further assume that the remaining tilt can be fully explained by the local normal vector of the platform at the positions of the OTs. Because of the platform deformation, the local normal vector will not be exactly parallel to the true pointing of the platform. That will cause a relative tilt of the OTs.

With Equation (\ref{eq:saddle}), the normal vector change contributed by the saddle-shape platform deformation can be written as:
\begin{eqnarray}
\delta n_{x,sadd}=-\frac{\partial dz}{\partial x}&=&-2A(y\sin 2\theta +x\cos 2\theta)\nonumber \\
                                     &=&-2Ar\cos(\alpha-2\theta)   \nonumber \\
\delta n_{y,sadd}=-\frac{\partial dz}{\partial y}&=&2A(y\cos 2\theta -x\sin 2\theta)\nonumber \\
                                     &=&2Ar\sin(\alpha-2\theta), \label{eq:normal1}
\end{eqnarray}
where $r$ and $\alpha$ are the polar coordinate coefficients of the OT location.

With Equation (\ref{eq:model2}), $\delta n_{x,sadd}$ and $\delta n_{y,sadd}$ can be written as 
the combination of two saddle pattern contributions:
\begin{eqnarray}
\delta n_{x,sadd}&=&-2A_{1}r\cos(\alpha-2\theta_1)-2A_{2}r\cos(\alpha-2\theta_2)\nonumber \\
\delta n_{y,sadd}&=&2A_{1}r\sin(\alpha-2\theta_1)+2A_{2}r\sin(\alpha-2\theta_2). \label{eq:normal2}
\end{eqnarray}
$A_{1}$, $A_{2}$, $\phi_{1}$, and $\phi_{2}$ are assumed to be functions of elevation.

Considering the common $r=0.6$m of the two optical telescopes, we can derive the relative tilt $\Delta n_{x,sadd}$ and $\Delta n_{y,sadd}$, which can be attributed to platform deformation, between two optical telescopes as following.
\begin{eqnarray}
\Delta n_{x,sadd}&=&2A_{1}r[(\cos\Delta\alpha -1)\cos\beta_{1}-\sin\Delta\alpha\sin\beta_{1}] \nonumber \\
            & &+2A_{2}r[(\cos\Delta\alpha -1)\cos\beta_{2}-\sin\Delta\alpha\sin\beta_{2}] \nonumber \\
\Delta n_{y,sadd}&=&2A_{1}r[(1-\cos\Delta\alpha)\sin\beta_{1}-\sin\Delta\alpha\cos\beta_{1}] \nonumber \\
            & &+2A_{2}r[(1-\cos\Delta\alpha)\sin\beta_{2}-\sin\Delta\alpha\cos\beta_{2}], \label{eq:normaldiff}
\end{eqnarray}
where $\beta_1=\alpha_1-az-2\phi_1$, $\beta_2=\alpha_1+2az-2\phi_2$, and $\Delta\alpha=\alpha_2-\alpha_1$.
$\alpha_1$ and $\alpha_2$ are the angular positions of OT1 and OT2, respectively.

We can further put $\Delta\alpha=-2\pi/3$ into Equation (\ref{eq:normaldiff}) and get:
\begin{eqnarray}
\Delta n_{x,sadd}&=&2\sqrt{3}A_{1}r\cos(\beta_{1}-\frac{5\pi}{6})+2\sqrt{3}A_{2}r\cos(\beta_{2}-\frac{5\pi}{6}) \label{eq:normaldiffx} \\
\Delta n_{y,sadd}&=&2\sqrt{3}A_{1}r\cos(\beta_{1}-\frac{\pi}{3})+2\sqrt{3}A_{2}r\cos(\beta_{2}-\frac{\pi}{3}).\label{eq:normaldiffy}
\end{eqnarray}

Considering the definition of $\beta_1$ and $\beta_2$, we can see that the two terms in the right hand side of Equation (\ref{eq:normaldiffx}) and (\ref{eq:normaldiffy}) correspond to the $n=1$ mode and $n=2$ mode in Figure \ref{fig:OTft}. (Note: The measured pointing error differs from the tilt of the OT by a minus sign.) Therefore, we can use the amplitudes and phases of $n=1$ and $n=2$ modes to determine deformation model parameters $A_1$, $A_2$, $\phi_1$, and $\phi_2$ as functions of elevation. Furthermore, we can also derive these parameters at different $hexpols$ with different sets of pointing error data. In other words, we can predict the platform deformation using the model (Equation (\ref{eq:model2})) with given pointing coordinate and polarization angle of the platform. With the knowledge of the location of each antenna on the platform, we can also predict the vertical displacements of all of the 13 antennas. That will lead us to the crucial geometric delays and phase delays.

However, the $n=0$ mode in Figure \ref{fig:OTft} is still not explained by our saddle model.
Because this $n=0$ mode is independent of pointing azimuth, we should be able to see significant changes
of the residual as shown in Figure \ref{fig:defresline} at different elevations if 
this $n=0$ mode is caused by the global pattern of the platform deformations. In Figure \ref{fig:defresline}
we see small differences between the residual deformation at $el=40^{\circ}$ and $el=60^{\circ}$. This difference
is not large enough to explain the large $n=0$ mode shown in Figure \ref{fig:OTft}. Consequently,  we think
this $n=0$ mode might be mainly due to some local effect near the locations where the OTs were mounted. In this case, we do not need to further consider the $n=0$ mode while determining the global platform deformation pattern. It is shown in section \ref{subsec:svd} that the platform deformation corrections fit the observed phases of visibilities very well without taking into account the $n=0$ mode.


One can also combine more than two rotating saddle patterns by rewriting Equation \ref{eq:model2} as linear combination of rotating saddle pattern terms such as $A_{m}[(x^{2}-y^{2})\cos 2\theta_{m} +2xy\sin 2\theta_{m}]$, where $\theta_{m}=(m/2)\times Az+\phi_{m}$, and $m$ is a integer. Following the similar way from Equation \ref{eq:normal2} to Equation \ref{eq:normaldiffy}, one can see that the saddle pattern rotating with $\theta_{m}$ will corresponds to Fourier mode with $n=\mid m\mid$. Therefore, for a platform with verified saddle pattern deformation, one can use Fourier modes of relative pointing errors of different OTs to determine the parameters of each of the rotating saddle patterns. In the case of AMiBA-13, the Fourier modes with $n>2$ are negligible. This fact implies that the saddle patterns with larger $m$ are also negligible.

\section{Correction for Platform Deformation}\label{sec:corr}
Using the model parameters derived above at different elevations, one can reconstruct the platform deformation with given platform pointing using Equation (\ref{eq:model2}). By feeding the positions of the 13 antennas on the platform into the platform deformation model, the vertical displacement of each antenna $d_{i}$ can be obtained. In this section, we will describe how we verify the model-predicted $d_{i}$ with real radio observations,
and the way we correct the phase errors induced by platform deformations using the model above.
We will also present how much the performance of AMiBA-13 has been improved with this correction.
\begin{figure*}
\plotone{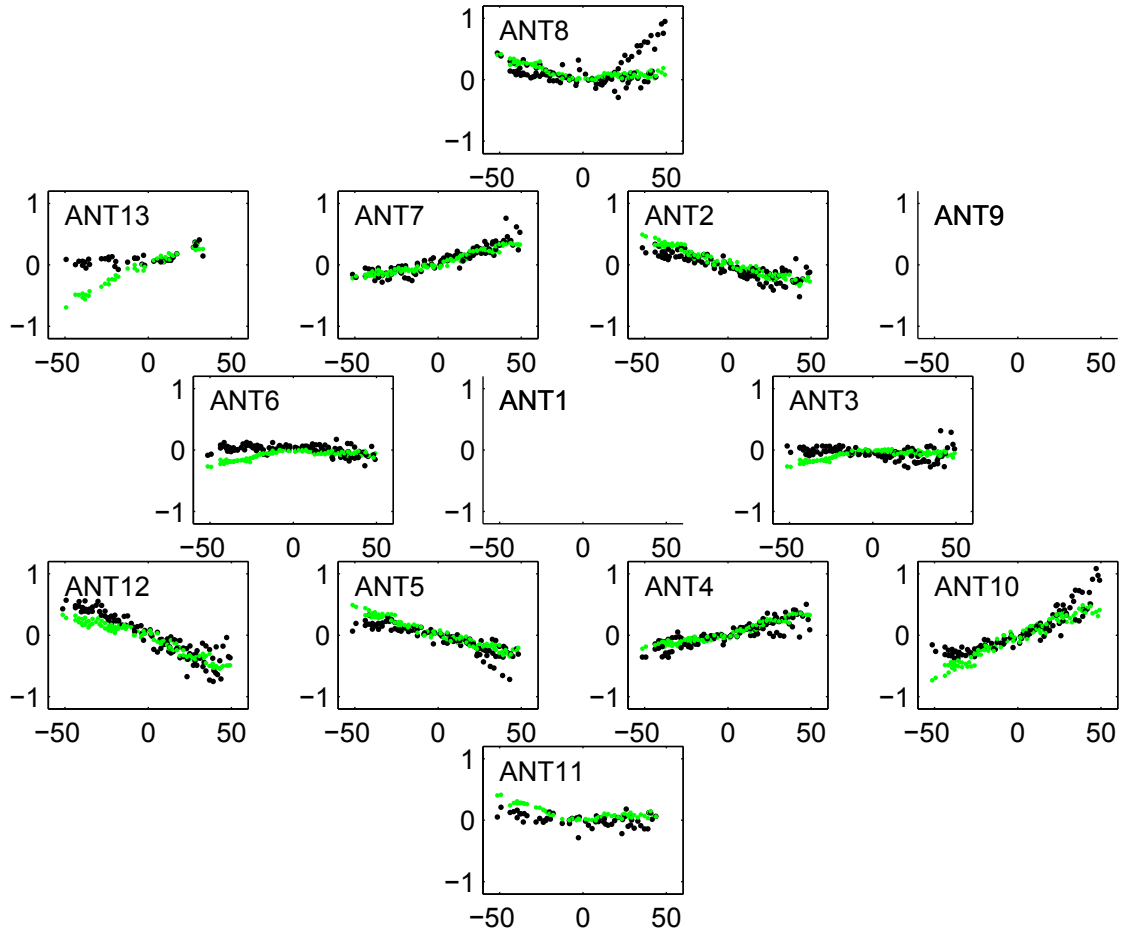}
\caption{Comparison between SVD solutions of vertical offsets (black dots) and predicted vertical offsets of the model (green dots). The SVD solutions were solved from the data of seven 6-hour long tracks on Jupiter. The x axis is hour angle in unit of degree, and the y axis is vertical offset in unit of mm. The vertical offsets of antenna 1 are set to be 0 and not shown here. The antenna 9 was temporarily off-line while doing the long tracks on Jupiter. \label{fig:svd}}
\end{figure*}
\subsection{Solving Deformations from Visibilities}\label{subsec:svd}
One important aspect of verification of the parametrized deformation
model is to check if the modeled geometric delays are consistent with those
measured from the visibility phases. For this purpose, we observed Jupiter,
which is considered a point source for AMiBA, with the two-patch subtraction method
 \citep{wu09} over several nights, covering as much as possible in
the range of hour angles. The uncertainty in the observed visibility phases were
dominated by systematic effects, while thermal noise could be ignored. The
major systematic effect was attributed to the band smearing effect when
transforming the limited 4-lag correlations to visibilities in two frequency
channels. Given a fixed transformation kernel, this effect sensitively depends
on the bandpass function of the receiver, intermediate frequency (IF)
electronics, and the individual analogue correlators \citep{lin09}. The
systematic uncertainty on the visibility phase thus has an antenna-based
contribution (from receivers and IFs) and a baseline-based contribution (from
correlators). One thing to remember is that the band smearing effect depends
on the total geometric delay (i.e. source offset and platform deformation),
and that it varies with telescope positions. 

The measured visibility phase between any antenna pair is related to their
geometric delays, instrumental delays, phase difference of local oscillator (LO)
signals, and the lag-to-visibility systematics. With a strong point source, its
pointing offset can easily be determined from the coherent phase distortion and
be corrected. Furthermore, Jupiter data taken each night were self-calibrated
with the visibility taken closest to its transit (highest elevation), leaving
only effects that vary with telescope positions:
\begin{equation}
 \phi_{ij}=k(\bar{d}_i-\bar{d}_j)+(\varphi_i-\varphi_j)+\psi_{ij},
 \label{eq:vis.phi}
\end{equation}
where $\phi_{ij}$ is the calibrated phase between antenna $i$ and $j$, $k$ is
the wave number, $\bar{d}_i$ is the vertical displacement for antenna $i$ relative to
when it is at near transit, $\varphi_i$ is the antenna-based relative systematic
uncertainty, and $\psi_{ij}$ is the baseline-based relative uncertainty.

Since the baseline-based systematics do not correlate with antennas, as an
approximation $\psi_{ij}$ can be dropped when solving Equation (\ref{eq:vis.phi}).
Equation (\ref{eq:vis.phi}) can be rewritten as
\begin{eqnarray}
 \phi_{\alpha} &=& (\delta_{\beta i} - \delta_{\beta j})_{\alpha}
\tilde{d}_{\beta}\nonumber\\
 &=& M_{\alpha \beta} \tilde{d}_{\beta},
\end{eqnarray}
where the index $\alpha$ refers to distinct pairs of antenna $i$ and $j$, $\beta$ refers to distinct antennas, $\delta_{\beta i}$ denotes the Kronecker delta, and $\tilde{d}_{\beta}$ is the relative vertical displacement of antenna $\beta$ including the antenna based systematics. In the case of AMiBA-13, we have index $i$, $j$, and $\beta$ ranging from $1$ to $13$, referring to 13 antennas, and index $\alpha$ ranging from $1$ to $78$, referring to 78 distinct antenna pairs.

The sparse matrix is clearly not square and is singular. We used the LAPACK
routine \emph{sgesvd} to perform the singular value decomposition (SVD) of the
sparse matrix $M_{\alpha\beta}$, zeroing singular values smaller than $10^{-6}$,
and then constructed the pseudo-inverse matrix $M^{-1}_{\beta \alpha}$ that was
applied to the measurement $\tilde{d}_{\alpha}$. However, since the equations
are based on differences of the variables, an arbitrary constant can be added to
the solutions. Here we define the vertical displacements of the central antenna 1 to be zero (Figure \ref{fig:svd}).

Once the solutions of the vertical displacements of each antenna from real radio observations have been obtained, one can compare the solutions with the predictions of the model above. Figure \ref{fig:svd} shows the comparison between model-predicted vertical displacements, and those solved from radio observation, for each antenna. As shown in Figure \ref{fig:svd}, the model predictions match the solved vertical displacements within a $0.2$mm range for most of the antennas and pointing coordinates. The good match between solved $\tilde{d}_{\beta}$ and the predictions of deformation model is a good evidence showing that other antenna-based systematics are secondary compared with the platform deformation.

\begin{figure*}
\plottwo{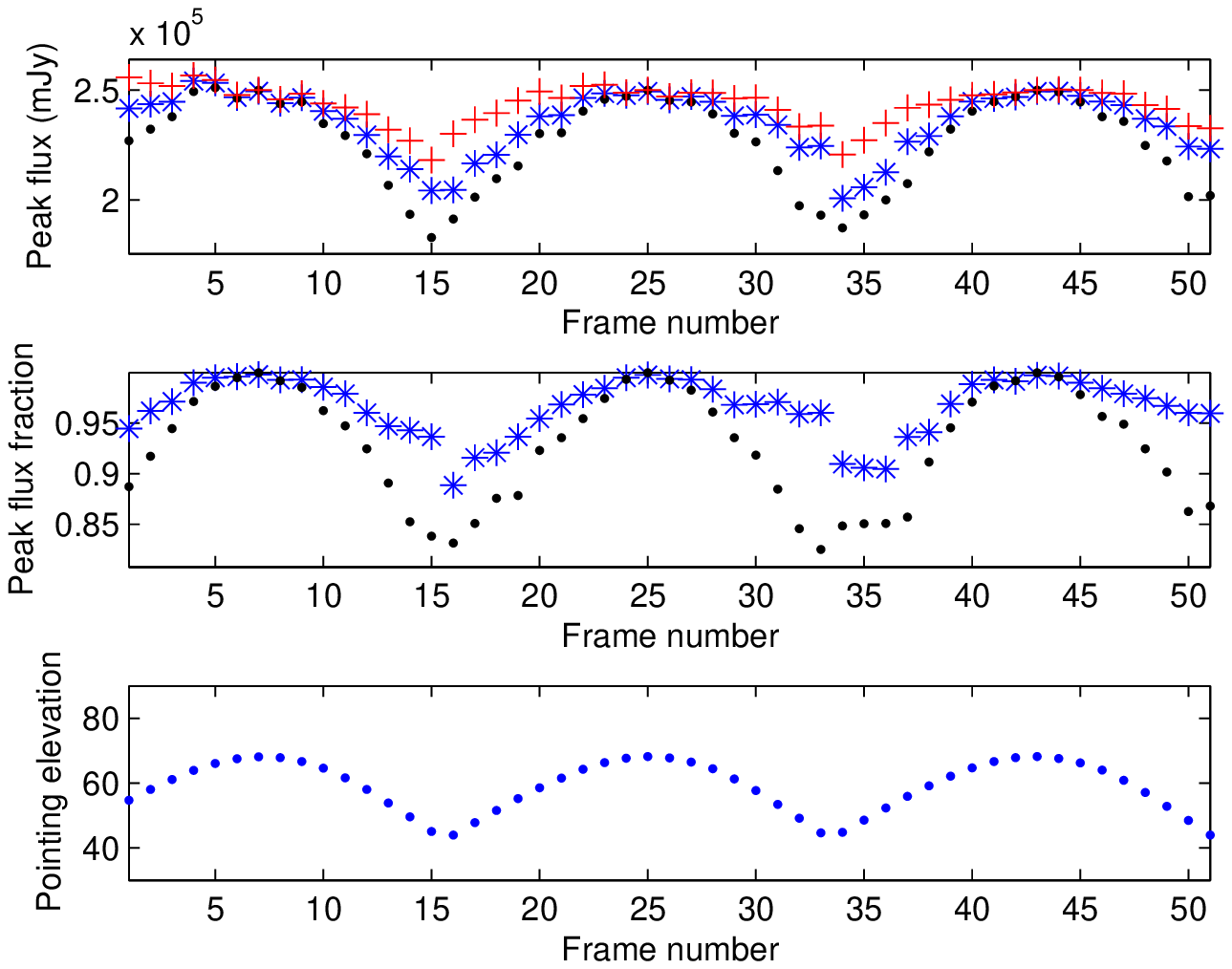}{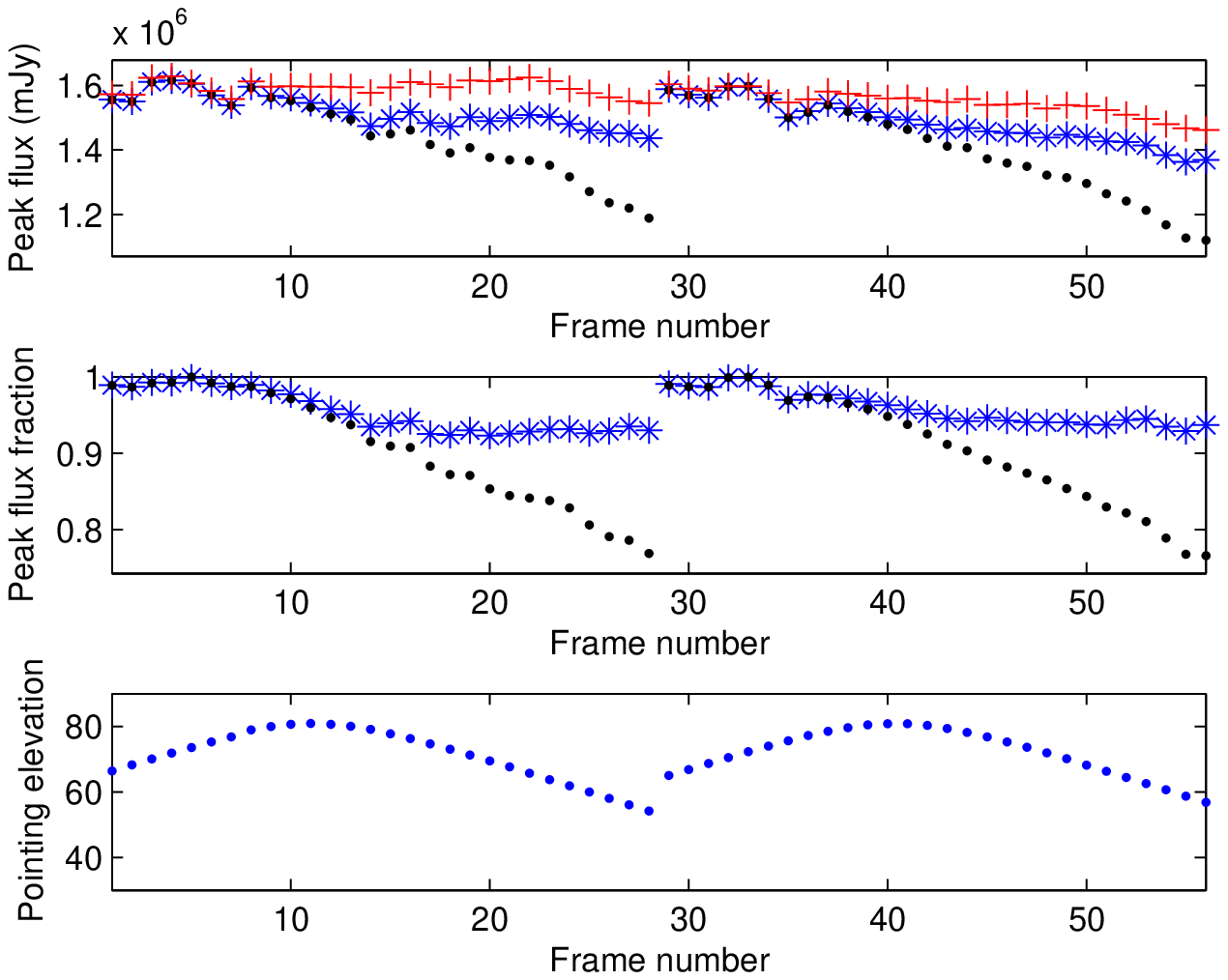}
\caption{The upper panels show the comparison between the peak values constructed from phase corrected visibilities 
(blue stars), uncorrected visibilities (black dots), and the simulated perfect phase correction (red crosses), of Saturn (left) and Jupiter (right) data. The middle panels show the corrected peaks and uncorrected peaks divided by perfect peaks. The lower panels show the pointing elevations of Saturn and Jupiter at each frame. The declination of Saturn and Jupiter was about $-2^{\circ}$ and $+11^{\circ}$, respectively. All the Saturn data were calibrated by the Saturn two-patches scans at the highest elevation every night. All the Jupiter data were calibrated by the Jupiter two-patches scans at $az\sim 119^{\circ}$ and $el\sim 72^{\circ}$ every night. The expected flux of Saturn and Jupiter was 238 Jy and 1600 Jy, respectively.
\label{fig:peaks}}
\end{figure*}

\subsection{Phase Correction of Visibilities}\label{subsec:phasecorr}
With the platform deformation model and pointing coordinate, we can predict the $d_{i}$ and $d_{j}$ required in Equation (\ref{eq:vis.phi}) to derive the geometric phase delay $\phi_{ij}$, which is induced by the platform deformation, for each pair of antennas. We simply multiply the calibrated visibilities $V_{ij,calibrated}$ by the exponential term $exp[i(\phi_{ij,target}-\phi_{ij,calibrator})]$ to correct the platform deformation induced phase error. $\phi_{ij,target}$ and $\phi_{ij,calibrator}$ are $\phi_{ij}$ derived with the coordinates of the observed target and calibration events, respectively.

We used Jupiter and Saturn data to quantify the improvement of the phase correction. Because the Jupiter and Saturn data are self-calibrated, we exactly know how strong the central flux should be with perfect correction. In order to do that, we take several 2-patches scans across the transit point on Jupiter and Saturn. Then we use one of the planet scans for each night as a calibrator to calibrate the data of the other scans, and construct the images with and without the platform deformation phase correction. For Saturn, we choose the scans closest to the transit point as the calibrators. For Jupiter, we choose scans with hour angles of about $-1$h to control the pointing error. Each two-patches scan was used to reconstruct one image with phase correction and one without correction. Figure \ref{fig:peaks} shows the comparison of the reconstructed central fluxes of Saturn with and without the phase corrections. We can see that the phase correction can recover about $30\%$ of flux lost of Saturn at $el\sim 45^{\circ}$. We can also recover about $50\%$ of flux lost of Jupiter at $el\sim 55^{\circ}$ (see Figure \ref{fig:peaks}).

We also constructed Saturn and Jupiter images using only amplitudes of the visibilities to simulate perfect phase correction. The peak values of these images are shown as red crosses in the upper panels of Figure \ref{fig:peaks}. By looking at these red crosses, we can find out that there will be still some flux lost even with perfect phase correction. One possible reason is likely to be the band smearing effect. Since AMiBA operates in a wide frequency band with two frequency channels, the incoming signal within a wide frequency range is integrated together. However, the phase delay due to the same geometrical delay changes at different frequency. Therefore, the signal integrated over a wide band is smeared by non-constant phase changes within the band. This smearing effect reduces the amplitude of observed visibilities. In order to evaluate the efficiency of the phase correction itself, we divided the peak flux values reconstructed with and without phase correction by the peak values derived with complete phase removal. The results are shown in the middle panels of Figure \ref{fig:peaks}. The phase correction can recover more than $50\%$ of flux lost due to phase error, restoring more than $90\%$ of the flux of the 'perfect phase correction', in most of the planet two-patches scans.

If the flux loss is due to band smearing, given that the amplitude of the platform deformation increases as the pointing elevation decreases, one can expect that simulated flux should decrease while the elevation becomes lower. The red crosses in the upper panels of Figure \ref{fig:peaks} show the expected trend. By simulating the perfect phase correction with Saturn and Jupiter data, we see that the remaining flux lost above elevation $40^{\circ}$ is smaller than $10\%$. Since the response across the entire bandwidth has not been measured in detail, it is difficult to estimate and correct the band smearing effect.


\subsection{Verification of Phase Correction with Radio Sources} \label{subsec:radio}
\begin{figure*}
\plotone{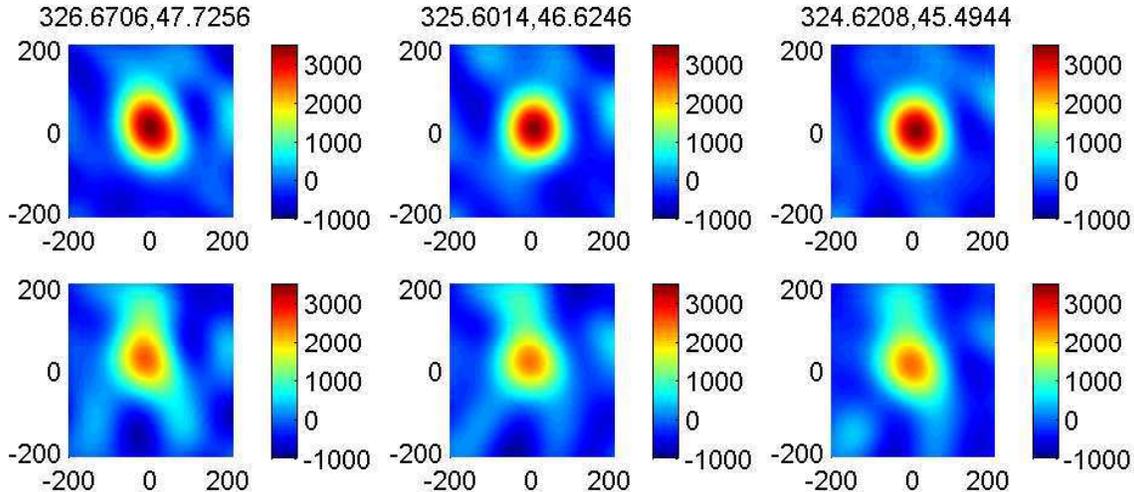}
\caption{An illustration of the improvement in image reconstruction contributed by phase correction. The target here is QSO B0355+508. The upper panels show the dirty maps reconstructed with visibilities after phase correction, while the lower panels show the results without phase correction. The angular unit here is arc-second, and the intensity unit is mJy. The azimuth and elevation of the platform pointings are shown in the image titles.
\label{fig:be_af}}
\end{figure*}

\begin{figure*}
\plottwo{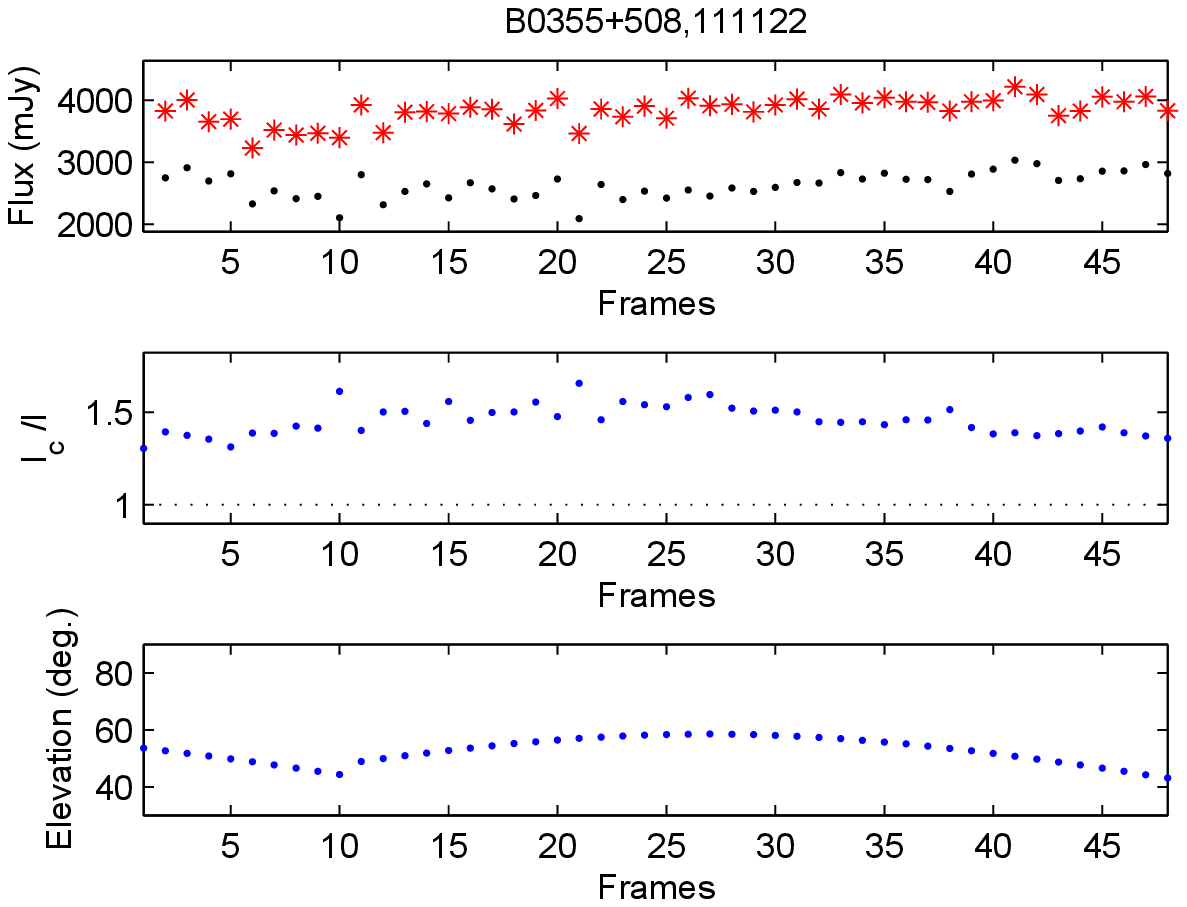}{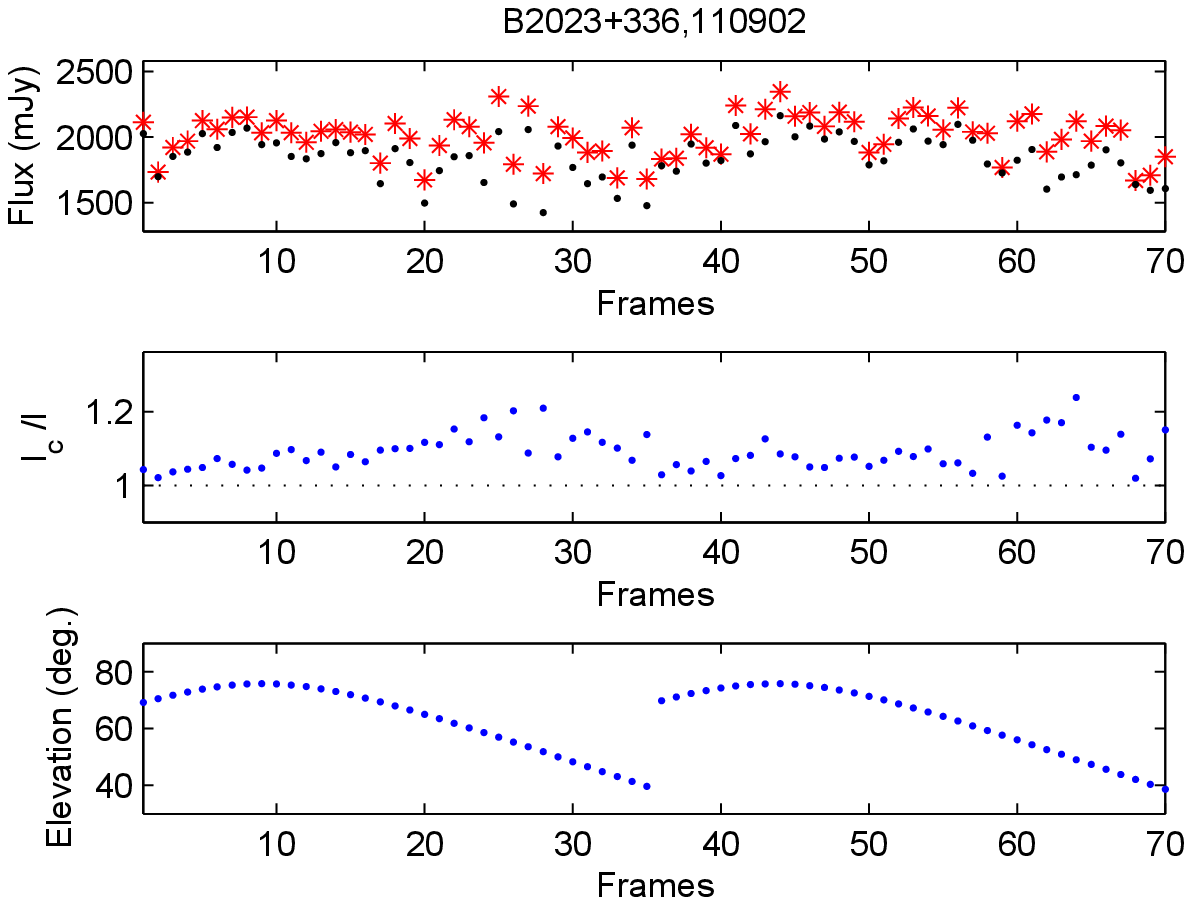}
\plottwo{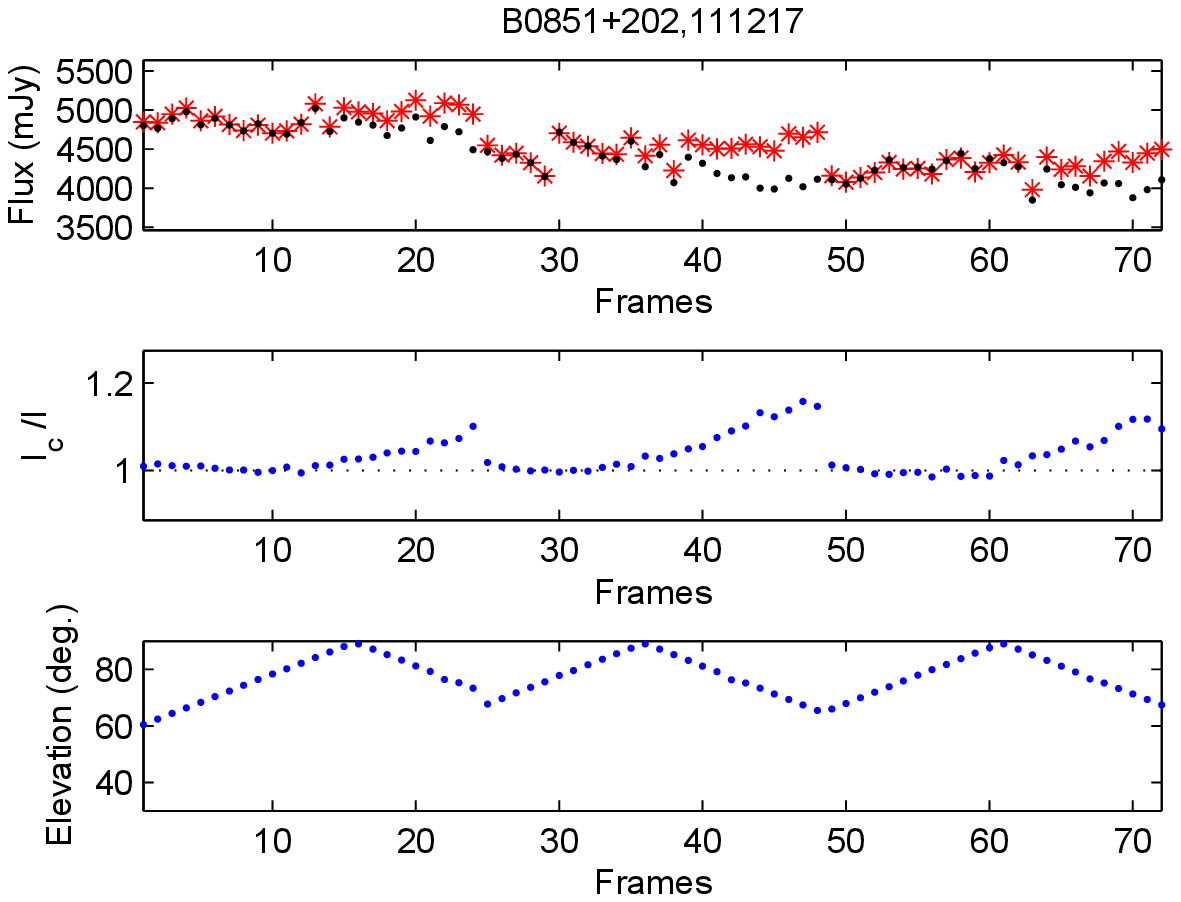}{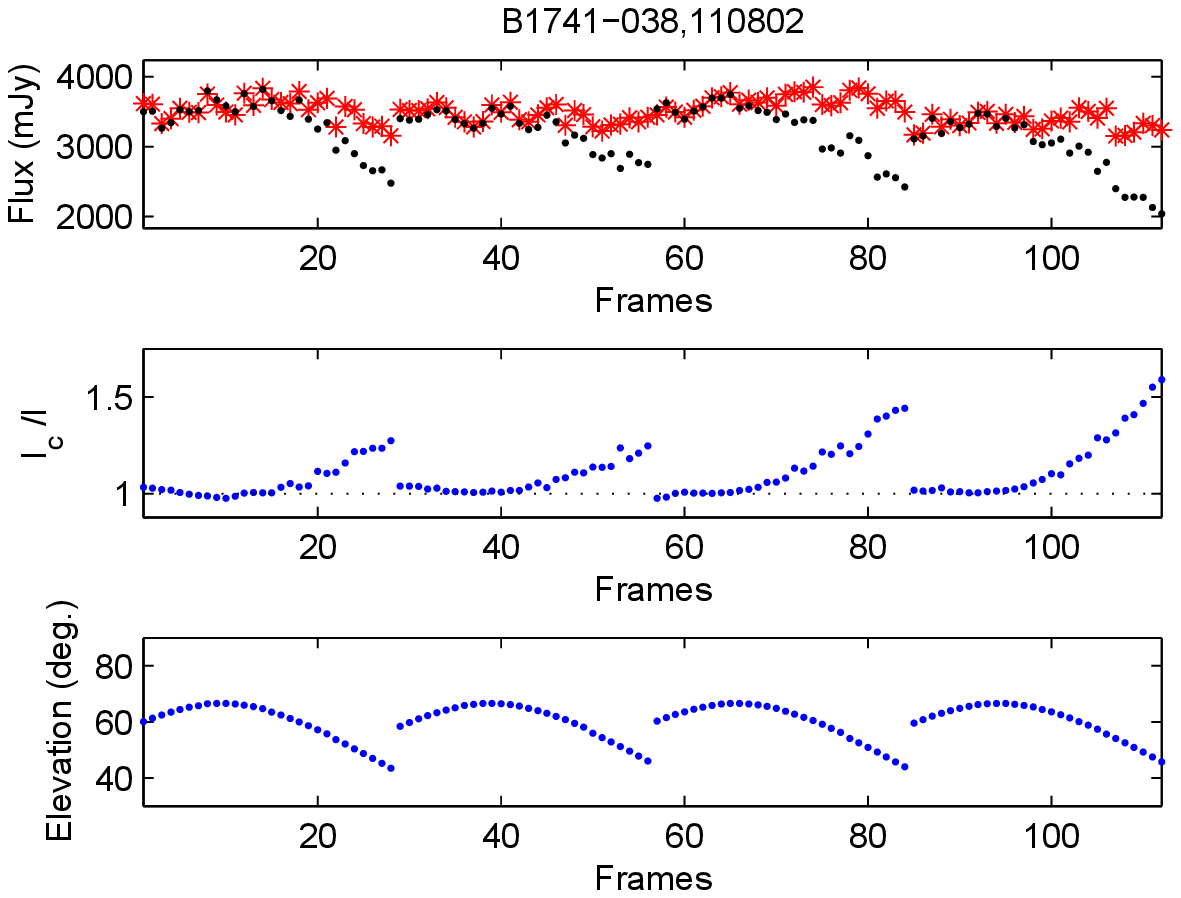}
\plottwo{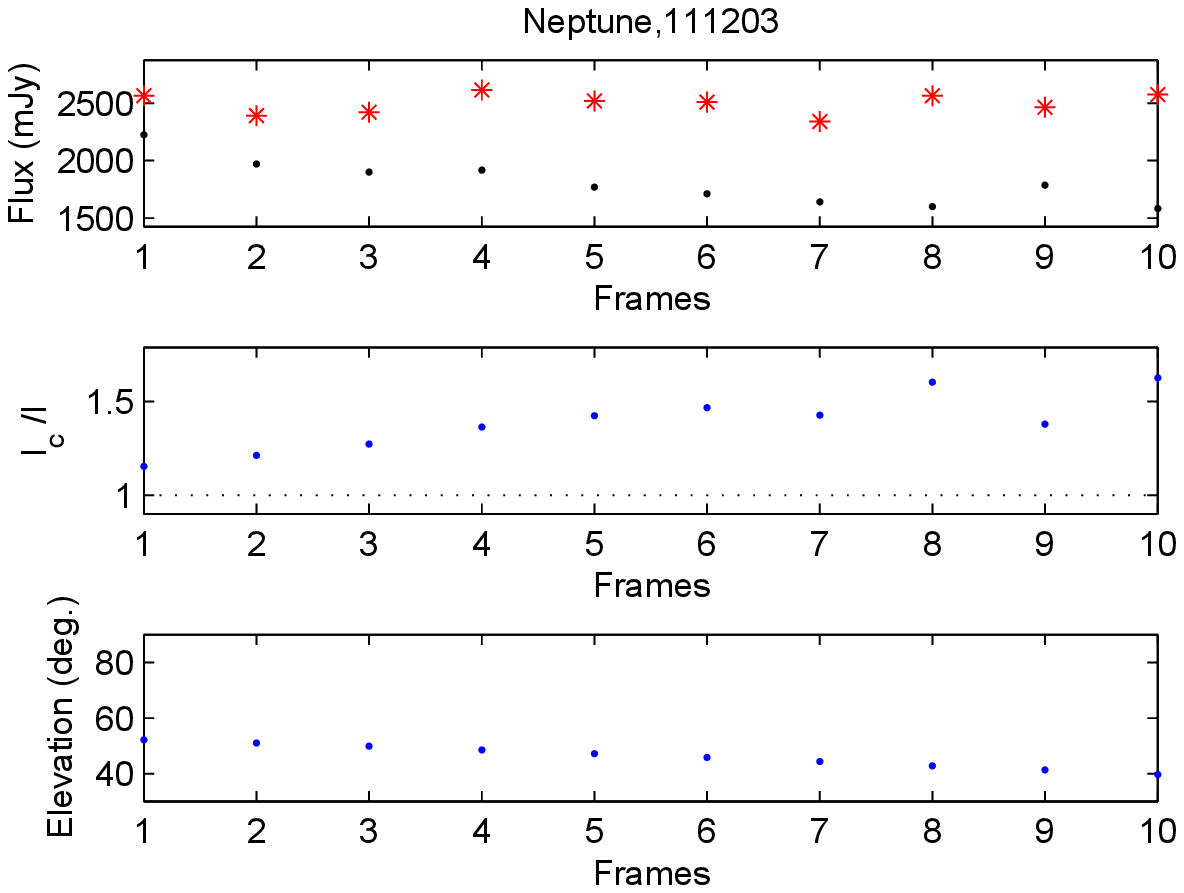}{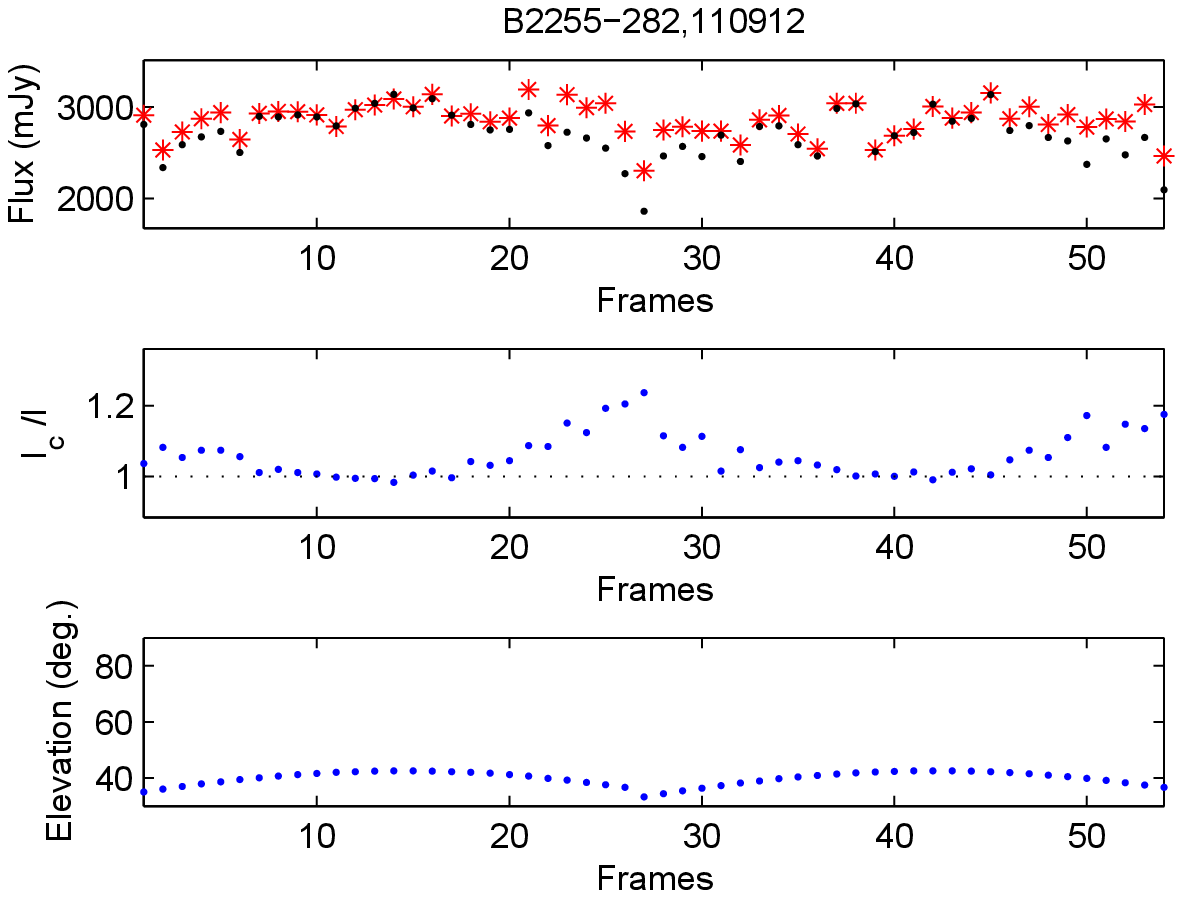}
\caption{The impact of phase correction on observed flux of radio sources and Neptune. The upper panels for each target show the observed fluxes. The black dots are uncorrected fluxes and the red stars are corrected ones. Each dot or star
shows the result from one 2-patch scan which takes 8 minutes. The middle panels show the ratios between corrected fluxes and the uncorrected ones. The lower panels show the elevations of each 2-patch scan.
\label{fig:peaks_radio}}
\end{figure*}

In the sections above, we corrected the phase delays and verified the efficiency of our correction with Jupiter and Saturn data. However, we can only verify the method within part of the pointing range with small changes of the declination of the planets. So we selected several radio sources with expected fluxes higher than 2 Jy at 90GHz from the ATNF catalogue \footnote[7]{http://www.narrabri.atnf.csiro.au/calibrators/}. We then did continuous two-patches scans on these radio sources. With one two-patches scan containing two 4-minutes patches, we can achieve S/N ratios higher than 30 for these strong radio sources. Other than the sources from ATNF catalogue, Neptune is also included in the samples of radio sources. The declinations of these targets are evenly spread between $+50^{\circ}$ and $-30^{\circ}$. The calibrator for this test is Jupiter. Most of the calibration scans were taken near $az=115^{\circ}$ and $el=73^{\circ}$, to control the pointing error induced by the calibration process. Figure \ref{fig:be_af} shows an example of the impact of phase correction on the image making. The peak fluxes of the images constructed with uncorrected and corrected visibilities are shown in Figure \ref{fig:peaks_radio}.

Assuming the phase correction is valid for these radio sources, we expect to see two features. The first one is an overall flux rise. The variances of the observed fluxes are also expected to be reduced after correction because phase correction is supposed to be able to remove the flux variances caused by deformation. As we can see in Figure \ref{fig:peaks_radio}, the phase correction increases the observed flux in almost all the 2-patch scans. The impacts of phase correction are more significant at low elevations as we expected. The variances of the flux between different scans on the same target are also reduced after phase correction. 

The phase corrections are expected to be smaller while the pointing of the observations are close to the calibration events. So one can also expect to see smaller flux changes after correction at pointing coordinates closer to the calibration events. Since most of the calibration events are in the eastern part of the sky, we can expect a stronger impact of phase correction in the western part of the sky as what we see in real data.

As proven to be valid on point radio sources, the phase correction method is applied to galaxy cluster data of AMiBA-13 in \cite{lin2013}.

\section{Discussions and Conclusion}\label{sec:discuss}
The deformation of the AMiBA platform is negligible in the AMiBA-7 configuration, but becomes a critical component of image reconstruction in the AMiBA-13 configuration. We present a new way to cope with the phase delays induced by the platform deformation.
We first use photogrammetry data taken at a few platform pointings to derive a generic platform deformation model. Then we use the difference of the optical pointing errors measured by two platform-mounted optical telescopes to estimate the deformation parameters within the entire pointing parameter space of the AMiBA platform. With the platform deformation modeling and phase correction method, we can control the flux loss to be less than $10\%$ for most of the platform pointings. 

The major part of the remaining flux loss is likely to be the result of a band smearing effect which is difficult to quantify exactly due to the low frequency resolution of the current AMiBA. With known response functions of each receiver and each baseline, it would be possible to further correct the band smearing effect with the platform deformation model. However, this part is currently not yet developed.

The saddle pattern deformation of AMiBA platform has been confirmed with both AMiBA-7 and AMiBA-13 configuration. Photogrammetry tests also show that changing the loading on the platform does not alter the saddle pattern deformation of it. Therefore, if the AMiBA antenna configuration changes in the future, one can perform multi-OT pointing test and use Equation (\ref{eq:normal1}) to check the saddle pattern. If saddle pattern is verified with multi-OT test, Fourier transformation of relative pointing errors between OT pairs can be done in the same way as described in Section \ref{subsec:ots}. The azimuthal dependence of phases of saddle patterns can be determined by different Fourier modes of $g_x$ and $g_y$, as described in Section \ref{subsec:ots}. If the multi-OT test results are not consistent to Equation (\ref{eq:normal1}), more photogrammetry might be needed to determine the new platform deformation pattern. The comparison between three or more OTs can also help to determine the origin of the Fourier mode with $n=0$. The $n=0$ mode is likely to be related to some local effects on or near a specified OT if $n=0$ mode is only seen in the OT pairs contain it.

The deformations of other instruments are not guaranteed to be like saddle pattern. Therefore, while applying the idea of this work onto another instrument, photogrammetry data should be taken at several well chosen mount position to understand the deformation model for that specified instrument. However, using optical telescopes to probe the model parameters for all of the telescope positions would be still valuable because photogrammetry tests for all of the positions could be avoid. That saves a lot of time.

Platform deformation can also be a critical issue for observation and image reconstruction as outlined here with other co-planar interferometers while the platform is not stiff enough. The same approach can be applied to other co-planar interferometers while coping with the platform deformation problem. It is also possible to apply this approach to investigate the deformation of single dish telescopes with several optical telescopes mounted on different locations on the primary mirror. The work of this paper also forms the basis for the AMiBA-13 data analysis and science results. Now we are using the method described in this paper to correct the galaxy cluster data of AMiBA-13. The corrections and analysis in the cluster data will be presented in \cite{lin2013}.

\acknowledgments
We thank the Ministry of Education, the National Science
Council (NSC), and the Academia Sinica, Taiwan, for their funding and supporting of the AMiBA project. 
YWL thanks M. Birkinshaw for helpful suggestions. The guiding, supporting, hard working, and helpful discussions of the entire AMiBA team are also acknowledged.

\end{document}